\documentclass[iop]{emulateapj-rtx4}

\shorttitle{XMM-Newton and Chandra study XB135}
\shortauthors{Barnard et al.}

\begin{document}


\title{XMM-Newton and Chandra observations of the M31 globular cluster black hole candidate XB135: a heavyweight contender cut down to size}


\author{R. Barnard and F. Primini and M. R.  Garcia}
\affil{Harvard-Smithsonian Center for Astrophysics (CFA), Cambridge MA 02138}
\and
\author{U. C. Kolb}
\affil{The Open University, Milton Keynes, UK}
\and
\author{S. S. Murray}
\affil{Johns Hopkins University, Baltimore, Maryland; CFA}


\begin{abstract}
CXOM31 J004252.030+413107.87 is one of the brightest X-ray sources within the D$_{25}$ region of M31, and associated with a globular cluster (GC) known as B135; we therefore call this X-ray source XB135. XB135 is a low mass X-ray binary (LMXB) that  apparently  exhibited hard state characteristics at 0.3--10 keV luminosities 4--6$\times 10^{38}$ erg s$^{-1}$, and the hard state is only observed below $\sim$10\% Eddington. If true, the accretor would be a high mass black hole (BH) ($\ga$50 $M_\odot$); such a BH may be formed from direct collapse of a metal-poor, high mass star, and the very low metalicity of B135 (0.015 $Z_\odot$) makes such a scenario plausible.  We have obtained new XMM-Newton and Chandra  HRC observations to shed light on the nature of this object. We find from the HRC observation that XB135 is a single point source located close to the center of B135. The new XMM-Newton spectrum is consistent with a rapidly spinning $\sim$10--20 $M_\odot$ BH in the steep power law or thermal dominant state, but inconsistent with the hard state that we previously assumed. We cannot formally reject three component emission models that have been associated with  high luminosity neutron star LMXBs (known as Z-sources); however, we prefer a BH accretor. We note that deeper observation of XB135 could discriminate against a neutron star accretor.
\end{abstract}




\section{Introduction}

One of the brightest X-ray objects in M31 is coincident with the globular cluster  (GC) known as B135, following the Revised Bologna Catalog (RBC) v 3.4 (Galleti et al., 2004, 2006, 2009). We refer to this X-ray source as XB135.  In our $\sim$13 year Chandra survey, it exhibited 0.3--10 keV luminosities $\sim$4--6$\times 10^{38}$ erg s$^{-1}$, assuming power law emission; the best fit photon index ($\Gamma$) was $\sim$1.8 throughout \citep{barnard14}. The high luminosity and hard spectrum of XB135 led us to propose that the accretor was a black hole (BH) with mass  $\sim$50 $M_\odot$, which is made possible by the very low metalicity of XB135 \citep[see][and references within]{barnard13}.
 We note that persistently bright BH binaries in GCs are consistent with  population synthesis theory for tidal capture of a main sequence star \citep[][although the companion might be disrupted]{kalogera04}, or with the formation of an ultracompact binary with a degenerate donor from a triple system \citep{ivanova10}.

We were awarded a 120 ks XMM-Newton observation of XB135, along with a 5 ks Chandra HRC observation. The XMM-Newton observation was designed to test the possibility that XB135 was formed in a low-metalicity environment. The purpose of the HRC observation was to confirm that XB135 consisted of a single point source that is consistent with being  located in B135;  XB135 had previously been observed at high off-axis angles, making its position uncertain. In this work, we fit the XMM-Newton spectra of XB135 with neutron star (NS) and BH XB  emission models; additionally, we  compare our XB135 spectrum with the XMM-Newton pn spectrum from the bright NS XB LMC X-2. We obtain an accurate position for XB135 from our Chandra HRC data.

\section{Observations and data analysis}

\subsection{XMM-Newton analysis}
XB135 was observed by XMM-Newton on 2012 June 26--27 (Obs ID 0690600401, PI R. Barnard). We analyzed the data using SAS version 13.0. A substantial portion of the observation was contaminated by background flaring. We screened out high background intervals in the pn instrument by creating a lightcurve  filtered with the expression ``(PATTERN==0)\&\&(FLAG==0)\&\&(PI in [10000:12000])'' and binned to 400 s; we rejected intervals with 10--12 keV rates $>$0.4 count s$^{-1}$. This resulted in 75 ks of good time.

Circular source and background regions on the pn image were chosen, and optimized by the software. Spectra were extracted from these regions using the expression ``(PATTERN$<$=4)\&\&(FLAG==0)'' and the good time interval. We did not use any data from the MOS cameras because they suffered significant pile up.
 In our proposal for this observation, we made a case for using the RGS spectra to help constrain the metalicity. However since much of the observation was lost to flaring, the good time RGS spectra were not of sufficient quality to constrain the fits.

We estimated our uncertainties  on spectral fits as follows. We first obtained the  best fits to the pn spectra. We then simulated 1000 sets of  spectra from this best fit model using the {\sc multifake} command in {\sc XSPEC}; variations in the simulated spectra were drawn from the statistical properties  of the observed spectra. Each simulated spectrum was modeled, obtaining the best fit parameter values and the 0.3--10 keV unabsorbed flux (2--20 keV flux in some cases, see below). Once all 1000 sets of spectra were analyzed, each parameter was sorted into ascending order, and the 1$\sigma$ confidence limits were obtained from the 16$^{th}$ and 84$^{th}$ percentile. The parameter values appear to be Gaussian distributed.

\subsection{Chandra analysis}
Our Chandra HRC-I observation was performed on 2013 February 24 (Obs ID 14400, PI R. Barnard). We registered this observation to the B band Field 5 M31 image provided by the Local Group Galaxy Survey \citep[LGGS,][]{massey06}, using three X-ray bright GCs, and following our procedure set out in \citet{barnard12}. This involved using the {\sc iraf} task {\sc imcentroid} to find the center coordinates for each GC   and its corresponding X-ray source, and using {\sc ccmap} to match the sky coordinates of the  GCs in the Field 5 image to the X-ray positions in the Chandra image. We used PC-IRAF Revision 2.14.1.

This procedure assumes that the X-ray sources are at the centers of their host GCs. Our assumption appears to be justified because the r.m.s. offset when aligning the X-ray and optical positions of 27 X-ray bright GCs in the M31 center was just 0.11$''$ in R.A. and 0.09$''$ in Decl. \citep{barnard12}, or  $\sim$0.3 ACIS pixels.

\section{Results}
\label{res}

\subsection{Locating XB135}

The native HRC pixel size is 0.13175$''$, and our image was binned by a factor 4 to $\sim$0.5$''$ pixels. The HRC-I observation revealed that XB135 is a single bright source, rather than a blend of multiple sources. Using the {\sc iraf} tool {\sc imcentroid}, we found the X-ray centroid with 1$\sigma$ position uncertainties of 0.07 image pixels in RA and Decl., i.e.   0.04$''$.


\begin{figure}
\epsscale{1.1}
\plotone{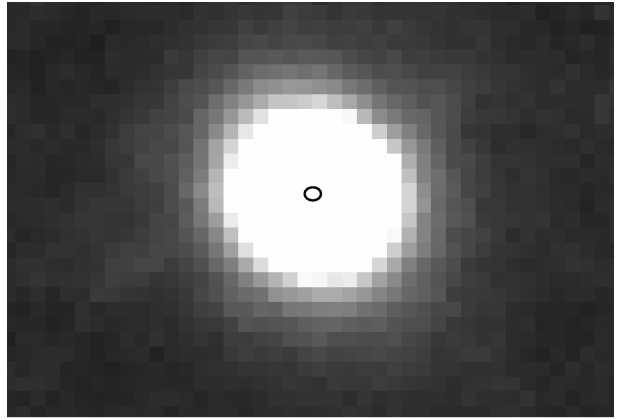}
\caption{B band image of the GC B135, from  Field 5 of the LGGS survey of M31 \citep{massey06}, overlaid with an ellipse that shows the 3$\sigma$ position uncertainty for XB135, using all 4 GCs for registration. The uncertainties from the 3 GC solution are entirely contained within this ellipse.  }\label{xbloc}
\end{figure}

We found a total of four X-ray bright GCs in the field-of-view: B045, B116, B135, and B091D. We initially registered the image using B045, B116, and B091D; this placed XB135 at the center of the GC B135, but the uncertainty in registration was negligible, perhaps due to our using the minimum number of sources required for the transformation. This resulted in a position of 00:42:51.985 +41:31:08.32, 0.08$''$ from the GC center in the Field 5 image. 

Additionally, we tried re-registering the image including XB135, to estimate the uncertainty in registration with an extra source; the resulting r.m.s. deviations  were 0.03$''$ in R.A. and 0.003$''$ in Decl. The position obtained from this solution  was 00:42:51.982 +41:31:08.31, 0.04$''$ from the Field 5 GC center; the combined 1$\sigma$ uncertainties (centroid position and registration)  were 0.05$''$ in RA and 0.04$''$ in Decl. 

For either solution, XB135 is confined to the  central region of the GC at the 3$\sigma$ level; this is consistent with the expectation that the at least some BHs migrate to the center because of  mass segregation \citep{morscher13}, possibly all of them \citep{spitzer69}. Figure~\ref{xbloc} shows the GC B135 in the Field 5 B band image provided by \citet{massey06}, overlayed with an ellipse representing the 3$\sigma$ position uncertainties for XB135 using all 4 X-ray bright GCs; the 3$\sigma$ position uncertainties obtained from the 3 GC solution are entirely contained within this ellipse. 

We have confined a BHC to the center of a GC for the first time. We therefore calculated the probability that XB135 is an unassociated bright X-ray source that happens to be within 0.2$''$ of the GC center ($\sim$10\% of the GC radius). 
 The XMM-Newton pipeline found 8 sources with fluxes $>$1.4$\times 10^{-12}$ erg s$^{-1}$ in our XMM-Newton observation, equivalent to luminosities $\ga 10^{38}$ erg s$^{-1}$ at the distance of M31. The EPIC pn field of view may be approximated by a circle with 14$'$ radius, hence the number density of bright X-ray sources is $\sim$4$\times 10^{-6}$ per square arc second. The probability of finding one of these bright X-ray sources coincidentally within 0.2$''$ of the center of any of the 80 confirmed GCs observed in the region \citep{peacock10}  is $\sim$4$\times 10^{-5}$. We therefore conclude that XB135 is indeed associated with the GC B135. We infer from this that the X-ray sources in B045, B116, and B091D are also located at the centers of their respective GCs, as suggested by our previous work.

\begin{figure}
\epsscale{1.3}
\plotone{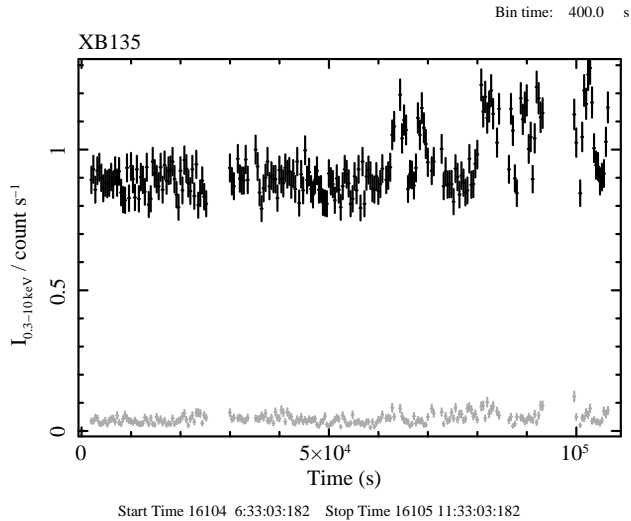}
\caption{Source (black) and background (gray) 0.3--10 keV pn lightcurves for XB135 from our 120 ks, 2012 June 26--27 XMM-Newton observation. Periods of intense background flares were removed. XB135 appears to have been somewhat variable, since it exhibited variation not seen in the background lightcurve; the r.m.s. variability was 10.1$\pm$0.6\%.  }\label{stlc}
\end{figure}

\subsection{Lightcurves}

\begin{figure}
\epsscale{1.2}
\plotone{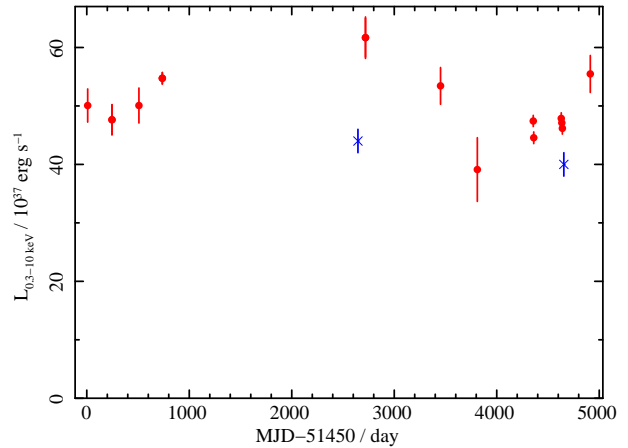}
\caption{Long term unabsorbed 0.3--10 keV luminosity lightcurve for XB135. Circles represent Chandra ACIS observations, while crosses represent XMM-Newton observations. }\label{ltlc}
\end{figure}

We present short-term and long-term lightcurves of XB135 in Figures~\ref{stlc} and \ref{ltlc} respectively. All luminosities assume a distance to M31 of 780 kpc \citep{stanek98}.

Figure~\ref{stlc} shows the 0.3--10 keV  pn intensity lightcurve of XB135 and the background for our 120 ks XMM-Newton observation. The lightcurve is binned to 400 s and periods of high background flaring are removed. XB135 exhibited some variability that is not apparent in the background lightcurve; the r.m.s. variability is 10.1$\pm$0.6\% on time-scales longer than 400 s. The hardness (2.0--10 keV intensity divided by 0.3--2.0 keV intensity) was consistent with being constant ($\chi^2$/dof = 253/216), from which we infer that this variation is energy-independent and likely due to the stochastic variations in accretion rate. 

Figure~\ref{ltlc} shows a 0.3--10 keV, unabsorbed luminosity lightcurve of XB135, using Chandra/ACIS and XMM-Newton observations, represented by circles and crosses respectively; the earlier XMM-Newton observation was analyzed in \citet{barnard08}.  The Chandra ACIS luminosities were derived from absorbed power law models \citep{barnard12}, while the XMM-Newton luminosities were derived from two-component models.
 The XMM-Newton observations appear to have  somewhat lower luminosities than the ACIS observations; it is possible that the ACIS luminosities are overestimated.
 However, the closest ACIS observation to our 120 ks XMM-Newton observation occurred 15 days earlier, and has a luminosity that is consistent within 3$\sigma$. Furthermore, the variation between the archival XMM-Newton observation and the nearest neighbor (a factor $\sim$1.4 over 73 days) is consistent with variability we have seen in bright X-ray binaries \citep{barnard12,barnard14}.


\subsection{Simple spectral modeling}

Since our XMM-Newton lightcurve  exhibited substantial variability, we accumulated  spectra using only the persistent emission (0.3-10 keV intensity $<$1 count s$^{-1}$); this spectrum contained 53500 counts, with 2\% coming from the  background. 

 We rebinned the spectrum to include at least 20 counts per bin. An absorbed power law fit to the spectrum ({\sc tbabs*po}) was rejected ($\chi^2$/dof = 1273/1033, good fit probability, g.f.p., = 4$\times 10^{-7}$). An absorbed disk blackbody fit ({\sc tbabs*diskbb}) was unsuccessful also ($\chi^2$/dof = 1185/1033, g.f.p. = 7$\times 10^{-4}$). 

\begin{figure}
\epsscale{1.1}
\plotone{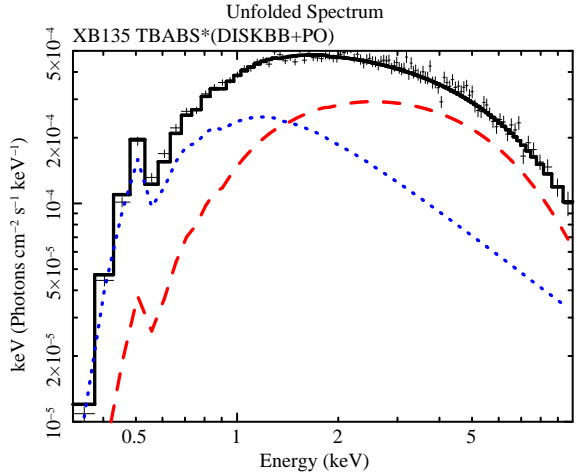}
\caption{Unfolded pn spectrum from our  XMM-Newton observation of XB135, with the best fit absorbed disk blackbody + power law  model. Dotted and dashed lines represent the disk blackbody and blackbody components respectively. The spectrum is multiplied by channel energy, in order to indicate the power at each energy. The data recieved further binning in this image for the sake of clarity.   
}\label{spec}
\end{figure}

Fitting a disk blackbody + power law yielded absorption ($N_{\rm H}$) = 2.7$\pm$0.3$\times 10^{21}$ atom cm$^{-2}$, k$T_{\rm DBB}$ = 2.15$\pm$0.08 keV, and $\Gamma$ = 2.2$\pm$0.3; all uncertainties are quoted at the 1$\sigma$ level; $\chi^2$/dof = 984/1031 (g.f.p. = 0.85). The 0.3--10 keV luminosity was 4.1$\pm$0.3$\times 10^{38}$ erg s$^{-1}$.  We present this fit to the unfolded pn spectrum in Figure~\ref{spec}. The 2--20 keV disk contribution to the flux was   69$\pm$11\% (1$\sigma$),  meaning that XB135 was not in a hard state (c.f. Remillard \& McClintock, 2006).  Instead, our spectrum of XB135 was consistent with either the thermally dominated state or the steep power law  (SPL) state.
 The inner disk temperature is substantially higher than for most Galactic BH XBs; however GRS 1915+105 has exhibited similar spectra, because the high BH spin results in a smaller ISCO and increased energy output \citep{remillard06}.

We note that the SPL  state is only observed during outbursts of transient BH XBs in our own galaxy. However, the two persistently bright,  dynamically-confirmed black hole + Wolf-Rayet binaries IC10 X-1 and NGC300 X-1 both exhibit emisson spectra consistent with the SPL state, as does the persistent BH XB LMC X-1; in \citet{barnard08} we suggested that such spectra are emitted by BH XBs with stable coronae, similar to the spectra observed from persistently bright NS XBs.

One possible difference between the spectrum of XB135 and the SPL state  is that the spectrum of XB135 exhibits significant curvature at energies above $\sim$2 keV. In this way it is similar to the high quality spectra observed from ultra-luminous X-ray sources (ULXs) studied by \citet{gladstone09}. Fitting our XB135 spectrum with a broken power law model ({\sc bknpower} in XSPEC) had $\Gamma$ change from 1.53$\pm$0.02 to 2.52$\pm$0.09 at 4.64$\pm$0.13 keV; these parameters are consistent with several ULXs analyzed by \citet{gladstone09}. They proposed that ULXs exist in a special ultra-luminous state, and this may be true because the luminosities of the ULXs are considerably  higher than that of XB135. However, it is possible that ULXs have similar spectra to other BH XBs (disk blackbody + power law) and the curved spectrum is caused by a dominant thermal component.

We note that some authors including  \citet{gladstone09}  consider  fits to be unphysical when the low energy spectrum is dominated by the Comptonized component. However, spectral analysis of ultra-luminous X-ray sources in NGC253 and the confirmed BH + Wolf-Rayet binary IC10 X-1 with XMM-Newton strongly suggests that this soft excess is real; fitting these spectra with the SIMPL convolution model for Comptonization created by \citet{steiner09} results in rejection by NGC253 ULX3 and IC10 X-1, and results in $\Delta \chi^2$ = +12 over the two component model fits with the same degrees of freedom for NGC253 ULX1 and NGC253 ULX2 \citep{barnard10}.

\citet{barnard10} proposed that this soft excess is due to an extended corona, and the evolution exhibited by IC10 X-1 during ``eclipse'' provided very strong evidence for a corona radius $\ga 10^{11}$ cm \citep{barnard14c}, consistent with with the empirical relation between 1--30 keV luminosity and corona radius observed in Galactic high inclination XBs \citep{church04}.  A coplanar, extended corona may feed from cool photons from the outer disk while heating the disk surface and stimulating production of cool photons \citep{haardt93}.


\subsection{BH emission models}

We decided to estimate the BH parameters for XB135 (assuming a BH accretor) by adopting a model that has been used to describe Galactic BH binaries.  Studies of the metalicity in the M31 bulge \citep{jablonka05} and part of the halo \citep{durrell01} show that the metalicity varies over 2 orders of magnitude in each region. While it is unlikely that the host GC B135 ($Z\sim 0.015 Z_\odot$) harbors much material, we obtained best fit spectra for models where the absorber metalicity  $Z$ = 0.1--1.0 $Z_\odot$ as well as $Z$ = 0.015.  To do this, we altered the metalicity with the {\sc abund} command in XSPEC.

The XSPEC model used was {WABS*TBABS*SIMPL*KERRBB}, described in the following paragraphs. 
For absorption, we included 7$\times 10^{20}$ atom cm$^{-2}$ at Solar metalicity to account for Galactic absorption (WABS); we also included an absorber with variable metalicity (TBABS, with the abundance controlled by the {\sc abund} command).

We  modelled the emission as SIMPL*KERRBB, describing the partially Comptonized emission from a disk around a spinning BH \citep[following][]{steiner09}. The variables for the emission model are: the fraction of disk emission that is Comptonized ($f_{\rm C}$), the BH spin parameter ($a$), the disk inclination ($i$), the BH mass ($M_{\rm BH}$), and  the accretion rate ($\dot{M}$, normalized to 10$^{18}$ g s$^{-1}$). The KERRBB model also requires a distance (assumed to be 780 kpc); the spectral hardening  factor, self-irradiation flag and limb darkening flag of the KERRBB model were kept at the default settings. For the Comptonized emission, we assumed $\Gamma$ = 2.4, as is typical for the steep power law state. In the end, the results proved rather insensitive to $\Gamma$.

Decreasing the M31 metalicity resulted in hotter spectra, translating as lower BH masses. The solar metalicity fit yielded the highest BH Mass: 21.8$\pm$0.8 $M_\odot$, with total $N_{\rm H}$ = 1.60$\pm$0.04$\times 10^{21}$ atom cm$^{-2}$, $f_{\rm C}$ = 0.57$\pm$0.03, $a$ = 0.998$\pm$0.005, $i$ = 65.7$\pm$0.7$^\circ$, $\dot{M}$ = 8.93$\pm$0.05$\times10^{17}$ g s$^{-1}$, and $\chi^2$/dof =  1008/1030. Assuming an absorbed metalicity of 0.015 $Z_\odot$ outside our Galaxy yielded a BH mass of 16$\pm$3 $M_\odot$, with $N_{\rm H}^{M31}$ = 4.9$\pm$0.4$\times 10^{21}$ atom cm$^2$, $f_{\rm C}$ = 0.34$\pm$0.10, $a$ = 0.95$\pm$0.02, $i$ = 78$\pm$3$^\circ$, $\dot{M}$ = 2.5$\pm$0.5$\times 10^{18}$ g s$^{-1}$ and $\chi^2$/dof = 977/1030. All uncertainties are quoted at the 1$\sigma$ level. 

We caution that these results should not be treated as  accurate, since we make assumptions in our modeling that may not apply. As such, we find that the models suggest a $\sim$10--20 $M_\odot$ BH  with spin $\sim$1, somewhat like the Galactic BH LMXB GRS 1915+105   which has a spin $>$0.98 \citep{mcclintock06}. They only considered spectra with luminosities $<$30\% Eddington when estimating the spin, to ensure that the disk is thin. Our Comptonized Kerr BH models yielded 0.3--10 keV luminosities $\sim$3$\times 10^{38}$ erg s$^{-1}$, $\sim$12--24\% Eddington for BH masses $\sim$10--20 $M_\odot$, which is in the range considered by \citet{mcclintock06}; our disk blackbody + power law fit yielded a luminosity $\sim$4$\times 10^{38}$, which is 16--32\% Eddington for a 10--20 $M_\odot$ BH. Unlike  GRS 1915+105,  XB135 appears to have a strong Comptonized component, which may complicate our interpretation.

We also note  that \citet{mcclintock06} identified a weakness in the {\sc kerrbb} model: the spectral hardening parameter (color temperature divided by effective temperature) is assumed to be a constant. Instead, they used a tabular model {\sc kerrbb2} where the hardening is computed.  However, the hardening factor for the range of parameters that we are interested in  is likely to be $\sim$1.7, the value assumed in our model (J. Steiner, private communication). Hence the differences between {\sc kerrbb} and {\sc kerrbb2} models are not expected to be great for XB135.

\subsection{Comparison with Galactic NS binaries}

Galactic NS XBs exhibit a wide range of behaviors; recently \citet{lin07, lin09, lin12} parameterised the full gamut by examining thousands of RXTE spectra from persistent and transient XBs exhibiting all spectral states using the same disk blackbody + blackbody emission model. They found that their model described all NS XB spectra well, with two exceptions: low accretion rate states where the  emission dominated by Comptonization \citep{lin07, lin09}; and the so-called Horizontal Branches of the highest luminosity NS XBs called Z-sources \citep{hv89} where a Comptonized component is required in addition to the disk blackbody and blackbody components \citep{lin09, lin12}.  \citet{lin10} successfully applied their double thermal model to Beppo-SAX and Suzaku spectra of a persistent NS XB, with energy ranges 1.0--150 keV and 1.2--40 keV respectively; hence the apparent success of the double thermal model is not confined to the RXTE band.   We note that this approach is not necessarily realistic, but the approach is valuable because it allows us to assess  the full gamut of NS LMXB behavior in a single parameter space.

We have identified 50 BHCs in M31 by comparing disk blackbody + blackbody fits to their spectra with the range of parameters exhibited by Galactic NS XBs \citep[see ][ and references within]{barnard14b}. 
For each BHC, we calculated the probability that the disk blackbody temperature, blackbody temperature, and the disk blackbody contribution were consistent with the NS XB parameter space ($P_{\rm DBB}$, $P_{\rm BB}$ and $P_{\rm f}$ respectively). Multiplying these probabilities together yielded the probability that the BHC was consistent with being a NS XB ($P_{\rm NS}$).
Since the only Galactic NS XBs to exhibit luminosities comparable with  XB135  are Z-sources, we compared our spectrum with the Z-source states studied by  \citet{lin09, lin12}; they found that $kT_{\rm DBB}$ $>$1.3 keV, $kT_{\rm BB}$ $>$2.2 keV, and $f_{\rm D}$ $>$0.66.

\subsubsection{Comparison with disk blackbody + blackbody emission models}
Lin et al. (2009; 2012) found that the majority of spectra were well described by disk blackbody + blackbody emission models. We created models with $Z$ = 0.015--1.0$Z_\odot$ using the XSPEC model WABS*TBABS*(CFLUX*DISKBB+CFLUX*BB); the fluxes were obtained in the 2--20 keV range for better comparison with the results of Lin et al. (2009; 2012).

As the metalicity of the M31 absorber increased, $kT_{\rm DBB}$, $kT_{\rm BB}$, and $f_{\rm D}$ systematically decreased; $kT_{\rm DBB}$ fell from 1.15$\pm$0.10 keV to 0.87$\pm$0.05 keV, $kT_{\rm BB}$ from 1.84$\pm$0.15 keV to 1.57$\pm$0.05 keV, and $f_{\rm D}$ from  0.41$\pm$0.08 to 0.26$\pm$0.03. $P_{\rm NS}$ ranged from 10$^{-5.1}$ to 10$^{-93}$.

\subsubsection{Comparison with disk blackbody + blackbody + power law emission models}
Lin et al. (2009; 2012) found that the Horizontal branch states of XTE J1701$−$462 and GX 17+2 required a Comptonized component, and found that the combined flux contribution of the disk blackbody and Comptonized component ($f_{\rm DBB+PL}$) was $>$76\% of the total. Hence we also made a series of models with WABS*TBABS*(CFLUX*DISKBB+CFLUX*BB+\\CFLUX*PO); we fixed the photon index of the Comptonized component to 2.4, as before. 

For these models $kT_{\rm DBB}$ and $kT_{\rm BB}$ systematically increased with increasing metalicity: $kT_{\rm DBB}$ from 1.14$\pm$0.10 keV to 1.4$\pm$0.4 keV, and $kT_{\rm BB}$ from 1.83$\pm$0.15 keV to 2.4$\pm$0.8 keV, with $f_{\rm DBB+PL}$ rising from 0.35$\pm$0.05 to 0.6$\pm$0.3. We find that for $Z$ $\ga$0.4, $P_{\rm NS}$ $>$0.01, and XB135 is consistent with NS XB spectra fitted with this emission model at the 3$\sigma$ level.  However, we note that the uncertainties in all parameters are significantly larger for this three-component emission model than for the previous two-component model; hence deeper observations of XB135, or with a broader energy range,  could potentially  separate XB135 from Galactic Z-sources for the three component model as well as the two component model.


\begin{figure*}
\epsscale{1.1}
\plotone{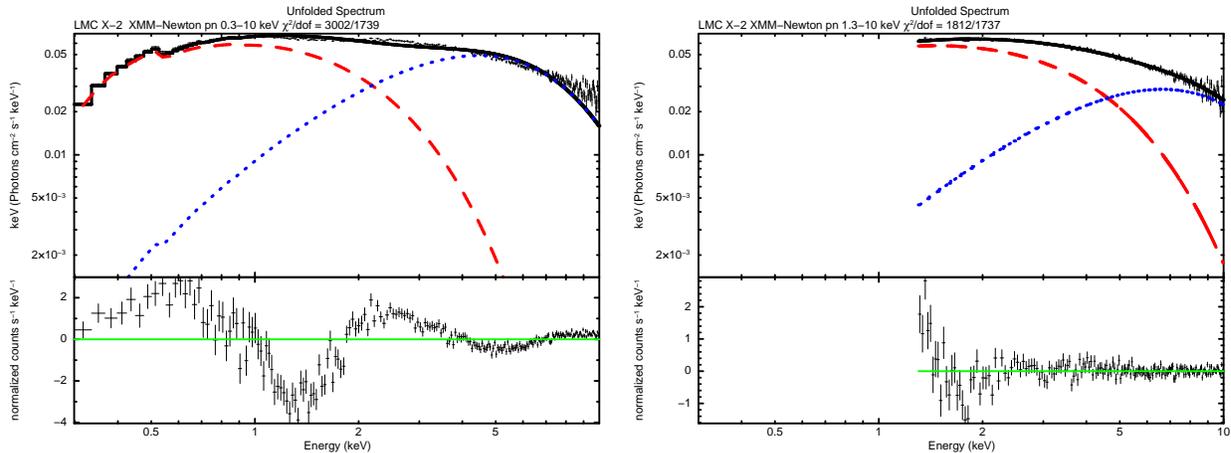}
\caption{Best fit disk blackbody + blackbody models for the 0.3--10 keV pn spectrum (left) and 1.3--10 keV pn  spectrum (right) from the Z-source LMC X-2. The dotted and dashed lines represent the disk blackbody and blackbody components respectively. The lower panel shows the $\chi^2$ residuals. For the left panel, we see considerable oscillation in the residuals, indicating a bad fit from the 0.3--10 keV spectrum. For the 1.3--10 keV fit, there are no systematic residuals, but some channels show large deviations that are probably due to uncertainties in calibration in this 750,000 count spectrum.  }\label{lmcx2}
\end{figure*}

\subsection{Comparison with an XMM-Newton observation of the Z-source  LMC X-2}

LMC X-2 is a NS XB residing in the Large Magellanic Cloud which exhibits spectral and timing behavior reminiscent of a Sco-like Z-source \citep{smale00,smale03}. It is also extremely bright for a NS XB, $\sim$0.5--2 $L_{\rm Edd}$ where $L_{\rm Edd}$ is the Eddington limit. It has a known counterpart, and a mass ratio $\le$0.4 \citep{cornelisse07}.

Unlike the Galactic Z-sources, LMC X-2 is sufficiently faint for observation with XMM-Newton; \citet{lavagetto08} present their analysis of the 2003 April 21 XMM-Newton observation of LMC X-2, which was conducted in Small Window mode with the Medium filter, allowing pn spectra that were free from pile-up. They used SAS version 7.0 and XSPEC version 11.3.2ad to analyze the pn and RGS spectra. After establishing that the background was stable, and that the spectrum varied little over time, \citet{lavagetto08} extracted source and background spectra from the whole observation. They found a large discrepancy between the pn and RGS spectra for energies below 1.3 keV, and attributed this to poor pn calibration for bright X-ray sources \citep{boirin03}; they therefore only used pn data above 1.3 keV.

\citet{lavagetto08} fitted  the pn and RGS spectra  with various models, including the disk blackbody + blackbody model adopted by Lin et al. (2007, 2009, 2012). Their best fits yielded  a disk blackbody temperature of 0.815$\pm$0.002 keV, and a blackbody temperature of 1.543$\pm$0.009 keV, which are significantly cooler than the parameters obtained from the Z-source spectra of  XTE J1701$-$462 and GX 17+2 \citep{lin09,lin12}  and remarkably similar to the parameters that we obtained for XB135. We therefore decided to investigate the XMM-Newton observation of LMC X-2 ourselves.

Using SAS version 13.0 and XSPEC version 12.8.1p, we modeled the pn spectrum above 1.3 keV using a disk blackbody + blackbody model (XSPEC model WABS*(DISKBB+BB) for consistency with Lin et al. (2007, 2009, 2012)). The spectrum contained $\sim$750,000 counts, with the source contributing $\sim$96\%. We binned the spectra to a minimum of 20 counts per bin. The absorption was unconstrained, so we adopted the best fit column density obtained by \citet{lavagetto08} for the disk blackbody + blackbody model, 4$\times 10^{20}$ atom cm$^{-2}$. 

Our best fit disk blackbody and blackbody temperatures for LMC X-2 were 1.36$\pm$0.04 keV and 2.42$\pm$0.08 keV respectively with $\chi^2$/dof = 1812/1737, very different to the results obtained by \citet{lavagetto08} and consistent with the parameters observed in Galactic Z-sources \citep{lin09,lin12}. When the temperatures were frozen at the values obtained by \citet{lavagetto08}, $\chi^2$/dof = 3002/1739. 

We present our best fit disk blackbody + blackbody fits to 0.3--10 keV  and 1.3--10 keV  pn spectra for LMC X-2 in the left and right panels of  Figure~\ref{lmcx2} respectively. We also present  residuals.  The 0.3--10 keV fit yielded unaccepatable results, with strong systematic variations in the residuals. The null hypothesis probability for the 1.3--10 keV fit (i.e. that differences between model and data are due to random variations in the spectrum) is 0.10, making this an acceptable fit even though $\chi^2$ $\sim$5--10  for some channels; these are likely due to uncertainties in calibration, which are only significant for very bright sources.

The fact that our new parameters for LMC X-2 are consistent with the Galactic Z-sources suggests that the results obtained by  \citet{lavagetto08} were probably affected by calibration errors, especially since they reported the discrepency between RGS and pn fits below 1.3 keV yet chose to fit the RGS spectra. We note that \citet{lavagetto08}  reported $>$5400 degrees of freedom in their spectra, and most of these come from low-weighted RGS bins; hence it is possible that their RGS+pn fit would be rejected if they had only considered the pn data.

 Fitting an absorbed power law model to the 1.3--10 keV pn spectrum of LMC X-2 allows us to compare the general shape of its spectrum with that of XB135. We were unable to constrain the absorption due to the 1.3 keV cut-off; however, \citet{lavagetto08} found $N_{\rm H}$ to be $\sim$4--10$\times 10^{20}$ atom cm$^{-2}$, depending on the emission model. For $N_{\rm H}$ = 1.0$\times 10^{21}$, $\Gamma$ = 1.373$\pm$0.003, with $\chi^2$/dof = 4546/1739, and lower values of $N_{\rm H}$ yielded harder $\Gamma$; such a spectrum is  significantly harder than any BH spectra \citep{remillard06}. Similarly, fitting XB135 with the best fit temperatures of LMC X-2 results in $\chi^2$/dof = 999/1033, whereas the equivalent free fit yielded $\chi^2$/dof = 973/1031; F-testing shows that the probability that this improvement is due to random variations is 1.2$\times 10^{-6}$, and the spectra of XB135 and LMC X-2 differ at a $>$4.8$\sigma$ level. The spectrum of XB135 is significantly softer than that of LMC X-2 when observed with the same instrument, strengthening our case for XB135 containing a BH accretor.


\begin{figure*}
\epsscale{1.1}
\plotone{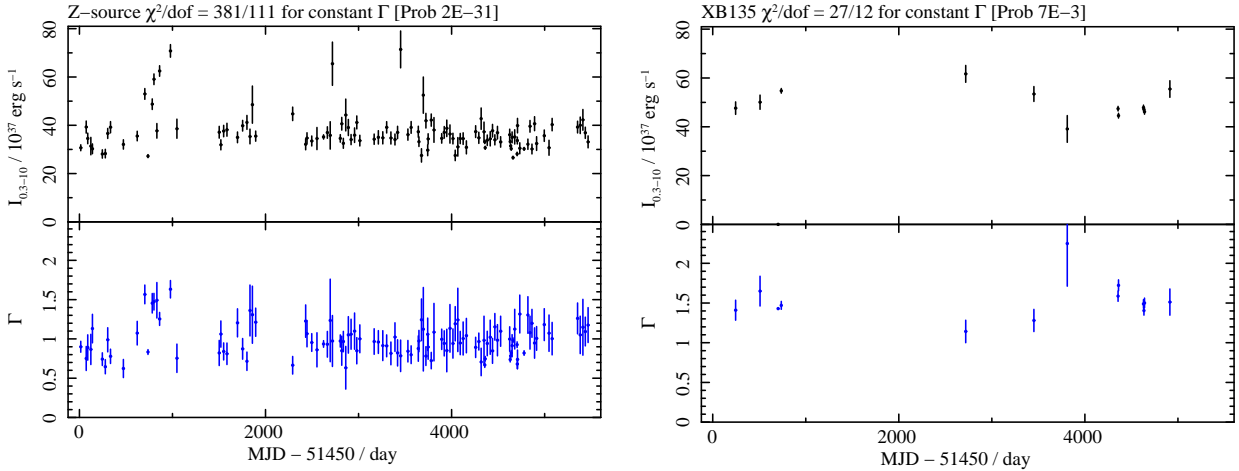}
\caption{Chandra ACIS 0.3--10 keV luminosity lightcurves (top) and spectral histories (bottom), for the M31 Z-source RX J0042.6+4115 (left) and XB135 (right), presented with equal scaling. We found the best fit absorbed power law model for each observation, and examined the spectral variability by fitting a line of constant photon index; XB135 is consistent with no spectral variation at the 3$\sigma$ level, while the Z-source RX J0042.6+4115 exhibited substantial spectral variability. }\label{lg}
\end{figure*}

\subsection{Comparison with M31 XBs}

While it is instructive to compare the X-ray populations of M31 and the Milky Way, this does not provide the full picture because the X-ray populations of these two galaxies are very different.

 \citet{stiele11} found 1897 X-ray sources within the M31 $D_{25}$ region at luminosities $>10^{35}$ erg s$^{-1}$; around 60\% of these sources were either classified only as Hard, or unclassified, and \citet{stiele11} expected these to be AGN. However, the Chandra survey of the M31 center (within $\sim$20$'$ or 4 kpc of M31*) revealed over 200 X-ray sources with luminosities $>$10$^{35}$ erg s$^{-1}$ that were missed by XMM-Newton, thanks to Chandra's lower background and superior PSF \citep{barnard14}. Furthermore, we identified $\sim$200 new X-ray binaries, and found strong evidence that the population of X-ray sources consistent with AGN contains a number of unidentified XBs. Hence there are more X-ray sources and fewer AGN within the M31 $D_{25}$ region than expected by \citet{stiele11}.

In \citet{barnard14b}, we present  analysis of 50 BHCs in M31, adding 13 transients and 2 GC XBs studied with XMM-Newton to our existing sample. By comparing the peristent and transient XB population within 6$'$ of M31$^*$, we found that $>$40\% of XBs $>$10$^{35}$ erg s$^{-1}$ are expected to contain BH primaries. Remarkably, BHC XBs represent  $>$90\% of those XBs with luminosities $>$10$^{38}$ erg s$^{-1}$; this is very different to the Milky Way, where the majority of XBs $>$10$^{38}$ erg s$^{-1}$ contain NS primaries. 

 Of the 15  BHCs associated with GCs, 12 exhibit disk blackbody + blackbody spectra that differ from NS spectra by $>$5$\sigma$ \citep{barnard14b}. While $\sim$30 M31 GCs contain X-ray sources that have exceeded 10$^{37}$ erg s$^{-1}$ at some point in the last 13 years, only one Galactic GC has ever been observed to exceed this luminosity. Hence the lack of confirmed BH XBs in Galactic GCs does not hinder our classification of XB 135 as a BHC. 

We know of two M31 X-ray sources that exceed 10$^{38}$ erg s$^{-1}$ in the 0.3--10 keV band, and most likely contain NS accretors. First is XB158, a high inclination X-ray binary that exhibits repeated intensity variations on a $\sim$2.8 hr period \citep{trudolyubov02} as well as a superorbital period of $\sim$5.8 days that was recently revealed by daily Swift observations over 30 days  (R. Barnard et al. 2015, ApJ submitted). Its 2002 XMM-Newton spectrum is consistent with disk blackbody + blackbody + power law fits to Galactic NS XBs; however, a simple power law fit yields $\Gamma$ 0.57$\pm$0.09, considerably harder than any BH spectrum (Barnard et al. 2015). Second is RX J0042.6+4115, which exhibited trimodal spectral evolution analagous to Galactic Z-sources \citep{barnard03}. Power law fits to our $\sim$100 Chandra ACIS observations of RX J0042.6+4115 yielded a mean $\Gamma$ of 0.856$\pm$0.008 \citep{barnard14}, although the spectrum varied considerably, as discussed below. While these power law emission models are clearly not realistic, they demonstrate that these  two  bright M31 NS XBs are significantly harder than XB135.

In Figure~\ref{lg} we compare the 0.3--10 keV lightcurves and spectral evolution from XB135 and the M31 Z-source candidate RX J00421.6+41152 over 13 years' worth of Chandra \citep[see][ for main survey results]{barnard14}. Z-sources evolved over time-scales of days to weeks \citep{muno02}, hence if XB135 were a Z-source, we should expect substantial spectral variability. We present the 0.3--10 keV lightcurve in the upper panel, and fit photon index for the best fit absorbed power law emission model in the lower panel; the scales are identical. We fitted each object with the best fit line of constant $\Gamma$. For XB135, $\chi^2$/dof = 27/12, with probability 7$\times 10^{-3}$, meaning that $\Gamma$ is consistent with being constant at the 3$\sigma$ level. For the Z-source RX J0042.6+4115, the best fit line of constant $\Gamma$ yielded $\chi^2$/dof = 381/112, with a probability of 2$\times 10^{-31}$. We note that the ACIS observations of RX J0042.6+4115 did result in pile-up; however, one would expect $\Gamma$ to decrease for a piled-up source as the intensity increased, yet  $\Gamma$ for RX J0042.6+4115 substantially increased, so the true spectral variation of RX J0042.6+115 was likely to be greater than observed. We present this as evidence in support of a BH accretor for XB135. We note that RX J0042.6+4115 exhibited flares during the 2002 XMM-Newton observation similar to those observed in Figure 2; however, the flares observed in RX J0042.6+15 were clearly energy dependent with higher amplitudes at higher energies \citep{barnard03}, in contrast with the energy-independent flares observed from XB135.

\section{Discussion and Conclusions}
We obtained new XMM-Newton and Chandra observations of the BH candidate associated with the M31 globular cluster B135. Previous results were suggestive of an extremely high BH mass of $\sim$50 $M_{\odot}$. The 120 ks XMM-Newton yielded $\sim$50,000 net source counts from intervals free from background flares and source variation. A high luminosity, hard state fit to such a high quality spectrum would have been extremely compelling evidence for such a massive accretor; unfortunately, such a hard state was ruled out. 
Fitting a black hole emission model to our XB135 spectrum suggests a black hole mass $\sim$10--20 $M_{\odot}$, considerably smaller than the $\sim$50 $M_\odot$ BH previously predicted. 

Our Chandra HRC observation provided the first observation of XB135 where the PSF was not greatly enlarged due to being observed at a high off-axis angle. We find that XB135 is a single point source, located at or very near the center of the GC B135 with a 3$\sigma$ position uncertainty of 0.15$''$ in R.A. and 0.12$''$ in Decl.

We prefer a BH primary for reasons stated earlier in the paper. We expect $>$90\% of M31 X-ray binaries with 0.3--10 keV luminosities $>$10$^{38}$ erg s$^{-1}$ to contain BH accretors \citep{barnard14b}, and $\sim$90\% of those very bright X-ray sources are located in GCs. Also, the 0.3--10 keV spectrum of XB135 is significantly softer than the similarly luminous Z-source LMC X-2 when studied with the same instrument (XMM-Newton pn), as well as RX J0042.6+4115 and XB158.  Furthermore,  XB135 exhibits less spectral variability than the M31 Z-source RX J0042.6+4115. Fitting the XB135 spectrum with a disk blackbody + power law model yields results that are consistent with the steep power law state when the uncertainties are considered; while the steep power law state is only exhibited by transient BH XBs in our own galaxy, persistent BH XBs IC10 X-1, NGC300 X-1, and LMC X-1 all exhibit spectra consistent with the steep power law state.

However, some aspects of the behavior observed from XB135 may contradict our BH assessment. We note that the best fit disk blackbody + power law parameters for XB135 are not obviously attributable to any canonical black hole state; however, the best fit disk blackbody + blackbody + power law model for XB135 is consistent with the Galactic Z-source studied by \citet{lin09,lin12}.  We also note that the type of flaring behavior observed from XB135 (see Fig. 2) is common for Z-sources, but not for BH XBs; however, the energy-independent flaring behavior from XB135 may be distinctly different from the energy-dependent flares observed from XMM-Newton observations of RX J0042.6+4115 \citep{barnard03}.

With the nature of the accretor in XB135 unclear, further observations are required. The uncertainties on the three component model would be constrained by either deeper observations, or by spectra with wider spectral ranges. Also, a better understanding of the metalicity of the absorbing material in M31  would also constrain the spectral parameters, and may aid the classification.

\section*{Acknowledgments}
We thank the anonymous referee for their constructive comments, which led to a significantly improved paper. This work was supported by the NASA ROSES-ADA grant NNX13AE79G, and the Chandra grant GO3-14028X.  This research has made use of  data from XMM-Newton, an ESA science mission with instruments and and contributions directly funded by ESA member states and the US (NASA).  We also include analysis of  proprietary data from the NASA Chandra X-ray observatory plus data obtained from the Chandra data archive, and software provided by the Chandra X-ray Center (CXC).

\acknowledgments

\email{aastex-help@aas.org}.



{\it Facilities:} \facility{XMM-Newton (pn)} \facility{CXO (HRC)}.

\end{document}